\documentstyle[aps,epsf,preprint]{revtex}
\begin{document}
\input epsf
\draft
\title{Deconfined Fermions but Confined Coherence?}
\author{S. P. Strong}
\address{
NEC Research Institute, 4 Independence Way,
Princeton, NJ, 08540, U.S.A.\\}

\author{David G. Clarke}
\address{IRC in Superconductivity and Cavendish Laboratory, 
University of Cambridge,\\
Cambridge, CB3 0HE, United Kingdom \\}
\date{June 11, 1996}
\maketitle
\begin{abstract}
The cuprate superconductors and certain organic
conductors exhibit transport which is qualitatively
anisotropic, yet at the same time other properties of
these materials strongly suggest the existence of
a Fermi surface and low energy excitations with
substantial free electron character.  
The former of these
features is very difficult to account for
if the material possesses three dimensional coherence,
while the latter is inconsistent with
a description based on a two dimensional
fixed point.  We therefore present a new proposal for these
materials in which they are categorized by a fixed point at which
transport in one direction is not renormalization
group irrelevant, but is intrinsically incoherent,
i.e. the incoherence is present in a pure system, at zero
temperature.
The defining property of such a state is
that single electron coherence is confined to lower dimensional
subspaces (planes or chains) so that it is impossible
to observe interference effects between histories which
involve electrons moving between these
subspaces.

\end{abstract}


\newpage

\section{Motivation}
\label{sec:motiv}

Interest in the physics of
strongly correlated, anisotropic materials has
undergone a resurgence since the discovery of the
cuprate superconductors. One of the central
questions in the theory of these materials
is the question of under what circumstances,
if any,
real bulk materials exhibit effective dimensionalities
other than three.
The question is certainly an important one
in the under and optimally doped cuprates
because they exhibit qualitatively anisotropic
transport.   For example, the temperature
dependence of the $c$-axis resistivity in under
and optimally doped
La$_{2-x}$Sr$_x$CuO$_4$ is nonmetallic at low temperatures
\cite{uchidadc}, while the $ab$ conductivity
remains metallic essentially all the way down to
$T_c$.  YBa$_2$Cu$_3$O$_{6+x}$ and Bi$_2$Sr$_2$CaCu$_2$O$_8$ 
exhibit activated $c$ axis conduction at low temperatures,
with $ab$ transport again remaining metallic \cite{ongdc}.
Meanwhile, the frequency dependent conductivity is
perhaps even more anomalous with 
no signs of anything remotely like a Drude peak 
with any appreciable weight in the 
$c$-axis conductivity in either underdoped
YBa$_2$Cu$_3$O$_{6+x}$ \cite{cooperac}
or La$_{2-x}$Sr$_x$CuO$_4$ \cite{uchidaac}.
In fact, $c$-axis conduction is rather generically
incoherent in these materials 
and has anomalous temperature dependence
in a range 
near to $T_c$.  This is difficult to understand
given that the scale of the expected
$c$-axis bandwidth from {\it ab initio} band structure
calculations is typically 2000K \cite{allen}.

Interaction effects can clearly reduce this
and one might attempt to explain some of the
anomalies in the $c$-axis conduction in 
terms of a reduced $c$-axis bandwidth.
Indeed, it is in principle possible for sufficiently
strong interactions to result in some 
non-Fermi liquid (NFL) in plane state with
respect to which the out of plane hopping would
be an irrelevant perturbation and this might
help to explain some of the transport
anomalies.  However, the criterion for
the renormalization group irrelevance of
the interplane hopping is that
\begin{equation}
\lim_{\Lambda \rightarrow 0}
\Lambda^2 \gg 
\lim_{\Lambda \rightarrow 0}
\int_0^{\Lambda} \int_0^{\omega} \int_0^{\omega}
d\omega d\omega_1 d\omega_2  \delta(\omega - \omega_1 -\omega_2)
\rho_e(k_F+0^+,\omega_1)
\rho_h(k_F+0^-,\omega_2)
\end{equation}
where $\rho_e(k_F+0^+,\omega_1)$ is the
spectral function for inserting an
electron with momentum just outside
the Fermi surface and
$\rho_h(k_F+0^-,\omega_2)$ is the
spectral function for creating a hole
with momentum
just inside the Fermi surface.
The requirement is clearly violated if 
either the electron
and hole spectral function
diverges at zero temperature and frequency
on the Fermi surface,
with the other either diverging 
or remaining finite. In fact,
the photoemission data
\cite{ding,shen}
show a sharp quasiparticle peak
at temperatures above $T_c$ and
it appears pathological to assume
that the peak does not narrow
arbitrarily 
on the Fermi surface as $T\rightarrow0$,
except for impurity effects
which will not affect the renormalization
group status of the hopping, $t_{\perp}$,
unless the impurity scattering rate is
large compared to the renormalized
hopping, $t_{\perp}^R$.  There is
also no theoretical reason for
a particle-hole anisotropy strong
enough to cause the hole
spectral function to vanish,
and it therefore appears highly unlikely
that $t_{\perp}$ is renormalization
group irrelevant.
It might be possible for 
$t_{\perp}^R$ to remain finite,
but small enough that
even a temperature of 40 Kelvin might
be taken to be high compared to
the $c$-axis bandwidth.  However,
in that case,
one expects to be able to describe the
conductivity in terms of hopping
conduction and one expects:
\begin{equation}
\sigma_c(\omega) \propto \frac{1}{\omega}
\sum_k \int d\epsilon \rho(k,\epsilon)
\rho(k,\epsilon + \omega)
\left( n_f(\epsilon) - n_f(\epsilon+ \omega) \right)
\end{equation}
This predicts a Drude peak at low frequency
given the form of the spectral function
observed in the photoemission, but no
such peak is seen.

This does not constitute an experimental proof
that something very peculiar is going on
regarding the dimensional crossover in the
cuprates, but it is certainly suggestive.
What seems to be required is a fixed point
at which $t_{\perp}^R$ is not zero,
or even particularly small, and yet conduction is
somehow very different from the usual,
three dimensional case.
It is the purpose of this work to make
a specific proposal for such a fixed point:
a state in which 
coherent transport 
is intrinsically
confined to the
planes, despite the fact that
the electrons themselves are not
\cite{prl,lees2}.
In this state a material would not
be three dimensional in the 
usual sense, but neither would it
be purely two dimensional.

\section{A Simple Model of Coupled Non-Fermi Liquids}


Unfortunately, our understanding of 2D NFL states
is not well enough developed to permit precise calculations to be made.
However, it is possible
to consider the analogous problem in one less
dimension, where we understand the
Luttinger liquid \cite{bosonization,haldjphysc}
behavior of interacting fermions.
We therefore consider the
problem of Luttinger liquids coupled by interliquid single particle
hopping as a potential paradigm for coupled NFL's \cite{general}. 

This problem is far from new: as early as 1974, in the context of the newly
discovered quasi-1D organic conductors, Gorkov and Dzyaloshinskii discussed
how various key properties of a 1D electron ``chain'' could be destroyed
by the presence of interchain hopping \cite{gorkov}. Many other authors have
since addressed the problem, using various techniques \cite{Bunch}.
To our knowledge, however, ours is the only approach which directly 
addresses the question of interliquid {\it coherence}. This question is of
crucial importance, for many of the other approaches {\em begin\/} with
an anisotropic 2D electron gas
(or its two chain analogue),
a state with manifestly coherent interliquid hopping,
upon which interactions are treated perturbatively. In those approaches
which do not begin with the anisotropic 2D electron gas, we believe
that,
while in some cases unreasonable approximations and/or errors have been made,
in general, these works have all correctly demonstrated the relevance of
$t_{\perp}$ but have simply not addressed the question of
its coherence.
In general, past workers have argued
that the flow away from $t_{\perp}=0$ should
lead to higher dimensional coherence
and,
for infinitely many chains, to a Fermi liquid
or to some other (CDW, SDW or BCS) known higher dimensional fixed point,
mainly because of the lack of an alternative proposal.
We are proposing 
the confinement of coherence as such an alternative.

The key construct needed in investigating the nature
of interliquid hopping is the electron spectral function, $\rho(k,\omega)$:
\begin{eqnarray}
\rho(k,\omega)&=&\theta(\omega)
\rho^+(k,\omega)+\theta(-\omega)\rho^-(k,-\omega)
\nonumber\\
&=&\sum_n\left\{\theta(\omega)
|\langle n_{N+1}|c_k^{\dag}|0_N\rangle|^2\delta(\omega-E_n^{N+1})
+\theta(-\omega)
|\langle n_{N-1}|c_k|0_N\rangle|^2\delta(\omega-E_n^{N-1})
\right\}
\nonumber
\end{eqnarray}
In a FL, $\rho(k,\omega)$ is dominated by a term which sharpens up to a 
$\delta$-function as $k\rightarrow k_F$. This term, of weight $Z_k\neq 0$, is 
the quasiparticle part of $\rho(k,\omega)$. The remainder of $\rho(k,\omega)$
is featureless so that, as far as low energy properties are concerned, 
only the quasiparticle part of $\rho(k,\omega)$
matters. 
If $\langle i,j \rangle$ label physically adjacent liquids, and $k$ in-liquid 
momenta, then
an interliquid hopping term of the form
\[
H_{\perp}=t_{\perp}\sum_{\langle
i,j\rangle,k}\left(c_{i,\sigma}^{\dag}(k)c_{j,\sigma}(k)
+{\rm h.c.\/}\right)
\]
will directly couple a quasiparticle state in one liquid with an
{\em energy degenerate\/} quasiparticle state
in the physically adjacent liquids.
In this case,
we should be doing
degenerate perturbation theory
and starting with 
interliquid Bloch states of precise interliquid 
momenta. 
An interliquid band will therefore form,
entailing a coherent interliquid
velocity and hence coherent interliquid transport.

In contrast, in a Luttinger liquid (or any NFL, by definition)
there are no Landau quasiparticles. The quasiparticle weight, $Z_k$, is
zero, but in a nontrivial way.
The Luttinger liquid spectral function
differs from that for a FL in that its singularities are power law in nature,
even at the Fermi surface. For the physically
most relevant case of spin-independent
electronic interactions, $\rho(k,\omega)$ has singularities at 
$\omega=\pm v_ck, v_sk$ determined by a single exponent $\alpha$,
and $v_c$ and $v_s$ denote velocities of propagation of charge and spin
currents. 
We believe that this  could potentially lead to
the confinement of 
coherence to the individual Luttinger liquids.
To see how this might come about, let us begin with
simple model within
which to discuss quantum coherence and incoherence: a two level
system (TLS) coupled to a dissipative bath.

\section{Incoherence and the Two Level System}
\label{sec:TLS}

To begin with we define, following Ref. \cite{TLS},
the two level system
model with the Hamiltonian:
\begin{equation}
\label{eq:TLS_ham}
H_{\rm TLS} =  \frac{1}{2} \Delta \sigma_x +
\sum_i  \left( \frac{1}{2} m_i \omega_i 
x_i^2  + \frac{1}{2m_i} p_i^2  \right)
+ \frac{1}{2}  \sigma_z \sum_i C_i x_i
\end{equation}
Here $C_i$ is the coupling to the $i$th oscillator,
and $m_i$, $\omega_i$, $x_i$ and $p_i$
are the mass, frequency, position 
and momentum of the $i$th oscillator,
respectively.

We restrict our discussion of the model to zero
temperature and the the so called 
ohmic regime \cite{TLS} where the spectral density of 
the bath is given by:
\begin{eqnarray}
\label{eq:ohmic}
J(\omega) &= &\frac{\pi}{2} \sum_i \frac{C_i}{m_i \omega_i} 
\delta(\omega-\omega_i) \\
\nonumber
 & = & 2 \pi~ \alpha~ \omega \exp(-\omega/\omega_c)
\end{eqnarray}
$\alpha$ is a positive constant measuring the strength
of the coupling to the bath and $\omega_c$ is a cutoff
frequency.

The model describes a single quantum mechanical
degree of freedom which can be in either of two states
and which is coupled to a bath of harmonic oscillators.
The environment, represented by the bath of
oscillators,
influences the tunneling
because the bath 
is sensitive to which of the states the spin
is in.  
Hereafter we will refer to the discrete degree of 
freedom as a ``spin'' for convenience. This is appropriate
since we are describing the spin with Pauli matrices.

The model provides the prototypical example
of a quantum to classical  crossover since, for
$C_i =0$ the model represents the quantum mechanics
of an isolated two state system, whereas for
sufficiently strong coupling to the environment
the dynamics of the spin,
if followed without reference
to the oscillator bath, are dissipative and
no quantum coherence effects are observable \cite{TLS}.
In fact, this is how one generally expects
classical behavior to emerge for macroscopic systems:
the macroscopic degrees of freedom exchange energy with
an enormous number of unobserved microscopic degrees
of freedom and therefore 
different histories are unable to maintain
a definite relative phase long enough for 
quantum interference effects
 to manifest themselves.

What sort of quantum interference effects do we expect to
be able to observe in the TLS for sufficiently weak coupling
to the environment?  Consider a model
where the coupling to the
environment vanishes, i.e. $\alpha=0$, and the system is
prepared 
in a state where the spin is in a $\sigma_z$
eigenstate.  The exact eigenstates of the spin are 
the $\sigma_x$ eigenstates which are split by an energy
$\Delta$ so that the initial state of the system is
a superposition of these two states 
of different energy with a definite
phase between the two states in the superposition.
Although not constant, this
relative phase remains
well defined indefinitely and 
therefore results in observable oscillations in
the expectation value of
$\sigma_z$, in fact 
(in units where $\hbar = 1$)
$\langle \sigma_z(t) \rangle =  \cos  \Delta t$.
In the TLS model 
the oscillations persist for a range of couplings
to the environment,
albeit with coupling dependent
damping of the oscillations.

To study these oscillations the standard
theory of the TLS focuses on a quantity called
$P^{(1)}(t)$ (hereafter simply $P(t)$)
which is the probability of finding the
system in the $\sigma_z=1$ state for
$t > 0$ for a system which has been prepared by clamping the
spin into the $\sigma_z =1$ state for all $t<0$,
allowing the oscillator bath to relax to equilibrium
in this configuration and finally releasing the spin
at $t=0$. 
Note that
calculation of $P(t)$ is equivalent to
determining $\langle\sigma_z(t)\rangle$,
since the two are simply related by 
$\langle\sigma_z(t)\rangle=2P(t)-1$.
$P(t)$ is an appropriate quantity to study for
questions about macroscopic quantum coherence since,
if the
spin represents a generic macroscopic quantum
degree of freedom which the experimenter can observe
and control, whereas the oscillators represent
microscopic degrees of freedom which are beyond
both control and observational capacities of the
experimenter,  it is exactly the sort of
preparation used in the definition of $P(t)$ which
is possible experimentally.  The signature of quantum coherence in
$P(t)$ will be the presence of oscillations
(damped or otherwise) in contrast to the incoherent
relaxation,
$P(t)\sim \frac{1}{2}(1+e^{- \Gamma t})$,
which results if the spin is sufficiently
strongly coupled to the environment, e.g.
when $\alpha > 1/2$ \cite{TLS}.
 
For further discussion of the TLS problem,
it is convenient to make a canonical transformation
on the original model, so that the coupling to the
oscillators is removed by taking:
\begin{equation}
H_{\rm TLS}^{\prime} = \hat{U} H_{\rm TLS} \hat{U}^{-1}
\end{equation}
where
\begin{equation}
\label{eq:xform}
\hat{U} = \exp \left(
-\frac{1}{2} \sigma_z \sum_i \frac{C_i}{m_i \omega_i^2}
\hat{p}_i \right)
\end{equation}
$\hat{p}_i$ is the momentum operator of the $i$th oscillator.
The new Hamiltonian takes the form:
\begin{equation}
\label{eq:TLS_ham_can}
H_{\rm TLS}^{\prime} = \frac{1}{2} \Delta(\sigma^+ e^{-i \Omega} + h.c.) +
H_{\rm oscillators}
\end{equation}
where $\Omega = \sum_i \frac{C_i}{m_i \omega_i^2} p_i$.
The tunneling operator
between the
two states has been replaced by an operator which creates and
destroys excitations of the oscillator bath,
as well as changing
the state of the spin.

In this formulation, $P(t)$ can be reinterpreted as the
probability of finding $\sigma_z(t)=1$ for a system in which
$\Delta$ is suddenly switched on at time $t=0$
with the system in the $\Delta = 0$ groundstate
with $\sigma_z =1$. The previous
definition in which the spin  was clamped in  the $\sigma_z=1$
eigenstate for all negative times and the oscillators were
allowed to adapt to the clamped state is equivalent.
Consider,
the two point correlation function of $\sigma^+ e^{-i \Omega}$,
which  obeys:
\begin{eqnarray}
\label{eq: Omega propagator}
\langle \sigma^+ e^{-i \Omega(t)} \sigma^- e^{i \Omega(0)}\rangle & = &
\exp \left\{-\int_0^{\infty} 
\frac{1 - e^{-i\omega t}}{\omega^2}
J(\omega)\right\}\nonumber\\
& = &
\exp \left\{-2  \alpha \int_0^{\infty} 
\frac{1 - e^{-i\omega t}}{\omega} 
e^{-\omega/\omega_c} \right\}\\
\nonumber & \sim &
e^{i \pi \alpha} (\omega_c t)^{-2 \alpha}
\end{eqnarray}
With the correlation function we can immediately construct the
spectral function of the operator $ e^{i \Omega}$
in the low energy, universal regime:
\begin{eqnarray}
\label{eq:TLSrho}
\rho_{\Omega}(\omega) & = & \sum_m |\langle m | e^{i \Omega} | GS \rangle |^2
\delta(\omega - E_m) \\
\nonumber & = & 
\Gamma^{-1}(2 \alpha)~\theta_+(\omega)~
\omega^{-1+2 \alpha} \omega_c^{-2 \alpha}
\exp(-\omega/\omega_c)
\end{eqnarray}
where $\{m\}$ is a complete set of oscillator eigenstates with
energies $E_m$ and $|GS\rangle$ is the oscillator ground state.
The  spectral function is normalized to integrate to unity
since $\langle e^{-i \Omega(t)} e^{i \Omega(t)} \rangle = 1$.

The short time approximation to $P(t)$ can be constructed 
straightforwardly using
the spectral function above and ordinary time dependent 
perturbation theory.  We find:
\begin{equation}
\label{eq:Pt1}
P(t) = 1 -  \frac{\Delta^2}{2}  \int d\omega \rho_{\Omega}(2\omega)
\frac{\sin^2(\omega t)}{\omega^2} + \cdots
\end{equation}
Notice that when $\alpha > 1$,
$\rho_{\Omega}(\omega) \sim \omega^{-1+2 \alpha}$ results in
an infrared convergent $P(t)$; in the limit $\Delta \rightarrow
0$, $P(t) \rightarrow 1$ for all $t$. This corresponds to
the irrelevance of $\Delta$ and the localization of the
spin predicted by Chakravarty 
and Bray and Moore \cite{chakrIsing} based on
a mapping of the TLS to the inverse squared Ising model.

Conversely,
for $\alpha \rightarrow 0$, $\rho_{\Omega}(\omega)\rightarrow\delta(\omega)$
and $P(t) = 1 - \frac{\Delta^2}{4} t^2 + ...$, in agreement with the
expansion of the exact result $P(t) = ( 1 +\cos\Delta t)/2$.
For $0 < \alpha < 1$ we are in a more complicated region.
Clearly the difference between $P(t)$  and $1$ grows to
order unity for any arbitrarily small $\Delta$ throughout
this region (the simply reflects the renormalization
group relevance of $\Delta$) and one would at first sight
be tempted to conclude that throughout this
region $P(t)$ would undergo damped oscillations with
a period approximately given by $t_{\rm osc}$, where $t_{\rm osc}$
satisfies:
\begin{equation}
\label{eq:tosc}
1 = \frac{\Delta^2}{2} \int d\omega \rho_{\Omega}(2\omega)
\frac{\sin^2(\omega t_{\rm osc})}{\omega^2} 
\end{equation}
One should be cautious, however, in view of the fact that
for $\alpha > 1/2$ the spectral function
for the tunneling operator is vanishing at low frequencies
and, at $\alpha = 1/2$, it is flat and
featureless out to the cutoff scale.  A flat spectral
function is equivalent to a featureless density of
states and is exactly the condition under which 
the Golden Rule approximation 
should be valid, implying incoherent decay without
any recurrence effects or oscillations. 
In fact, the exact solution of the TLS at the 
special value $\alpha = 1/2$ shows purely incoherent
relaxation in $P(t)$, and, while
the true behavior of $P(t)$ in the region $1/2 < \alpha <1$ is 
not
rigorously known \cite{TLS}, there are reasons for believing that
the behavior there is no more coherent.
The simple reason for this is that since,
in these cases the spectral function has even more
high energy weight, 
perturbation theory in $\Delta$ is even
less coherent than for $\alpha = 1/2$.
Incoherence is the direct result of the 
non-degeneracy of the perturbation theory in $\Delta$.
Let us discuss this important point in more detail.

The important
physical
effect of finite $\alpha$ is that there is a substantial contribution
to $P(t)$ from transitions to states with energies that are
larger than the putative renormalized oscillation frequency,
$\Delta_R \sim t_{\rm osc}^{-1}$ (see Eq. \ref{eq:tosc}) .
When the amount of weight in these transitions is larger
than the amount of weight in transitions to low energy
states, it no longer makes sense to consider the effects
of $\Delta$ to be coherent. Effectively, each change of state,
{\em i.e.\/} flipping of the spin,
is accompanied by the creation or annihilation of a sufficient
number of bosons in the environmental bath that the 
phase of that history is randomized compared to histories
with no spin flip.
Intuitively, {\em one has crossed over
from degenerate or nearly degenerate perturbation theory to
non-degenerate perturbation theory} (as opposed to the 
transition to irrelevant $\Delta$ where the long time
perturbation theory becomes convergent). 

To see that this picture does agree with the
known TLS results, let us calculate the
contribution
to $P(t)$ from transitions to states with 
various energies.
The low energy part
contributes to $P(t)$ an amount
\begin{eqnarray}
\delta P_{\rm low}(t) & = &  \frac{\Delta^2}{2}  \int_0^{1/t}
 d\omega \rho_{\Omega}(2\omega)
\frac{\sin^2(\omega t)}{\omega^2} + \cdots \nonumber \\
& \sim &
\frac{\Delta^2}{2^{2-2\alpha}} t^{2-2\alpha} \Gamma^{-1}(2 \alpha) ~ 
\frac{_1F_2(-1+\alpha;\frac{1}{2},\alpha;-1) -1}
{4 (1-\alpha) }
\end{eqnarray}
where $_1F_2$ is a generalized hypergeometric function
\cite{gradrhyz}.
This evaluates to $\frac{1}{4} \Delta^2 t^2$ at $\alpha=0$
and, for $\alpha = 1/2$, to
$\frac{1}{4}\Delta^2 t\,(2\,{\rm Si}(2)+\cos 2 -1)\approx 0.45 \Delta^2 t$
(${\rm Si}$ is the sine integral function).

The high energy part contributes
\begin{eqnarray}
\delta P_{\rm high}(t) & = &  \frac{\Delta^2}{2}  \int_{1/t}^{\infty}
 d\omega \rho_{\Omega}(2\omega)
\frac{\sin^2(\omega t)}{\omega^2} + \cdots \nonumber \\
&\sim&\frac{\Delta^2}{2^{2-2\alpha}} t^{2-2\alpha} \Gamma^{-1}(2 \alpha)
\left(\frac{1}{4 (1-\alpha)  } +
\frac{\Gamma(2\alpha)\cos\pi\alpha}{2^{2\alpha}(1-2\alpha)(1-\alpha)}
-\frac{_1F_2(-1+\alpha;\frac{1}{2},\alpha;-1)}
{4 (1-\alpha)  } \right)
\end{eqnarray}
For $\alpha \rightarrow 0$ the high energy part vanishes like
$\Gamma^{-1}(2 \alpha)$ (the prefactor is
$\frac{1}{4} \Delta^2 t^2 \left( \frac{3}{2} - \gamma -\ln2 +
\frac{1}{6}~_2F_3(1,1;2,5/2,3;-1) \right)
\approx 0.1 \Delta^2 t^2$), while for $\alpha =1/2$
the result is 
$\frac{1}{2\pi}\Delta^2 t\,(1+\pi-_1F_2(-\frac{1}{2};\frac{1}{2},\frac{1}{2};-1)
\approx 0.34 \Delta^2 t$,
comparable to the low energy contribution.

Clearly, the high energy contribution
is insignificant as $\alpha \rightarrow 0$ because of
the divergence of $\Gamma(2 \alpha)$ (the gamma function has
a simple pole at $0$) and,
for any arbitrary division into ``high'' and ``low'',
could always be made so by taking a
suitably small $\alpha$.
One therefore expects to find coherence in the
limit $\alpha \rightarrow 0$. On the other hand, for finite values
of $\alpha$ the high energy part can be as important as the low energy part,
depending upon our division into high and low energy integrals.
Using the qualitatively reasonable division above, we
see that for $\alpha = 1/2$ the two contributions
are in fact comparable.  
In fact, for any division scheme involving energy scales small
compared to the oscillator cutoff,
the high energy part must dominate for
some $\alpha < 1$, since, in 
the limit where $\alpha \rightarrow 1$,
the high energy part diverges logarithmically like
$\frac{\Delta^2}{2  \omega_c^{2}}
\ln(\omega_c t)$
while the low energy part is finite
and given by 
$\frac{\Delta^2}{2  \omega_c^{2}}
(\gamma + \ln(2) - {\rm Co}(2)) \approx 0.85~\frac{\Delta^2}{2 \omega_c^{2}}$,
where ${\rm Co}$ is the cosine
integral function and $\gamma$ is Euler's constant.
The high energy part can therefore
be made arbitrarily large compared
to the low energy part for any arbitrary  partition into
high and low energy pieces as we approach $\alpha = 1$.  
The dominance of the high energy part does not necessarily
imply that the quantum oscillations must cease entirely;
it could be that the oscillations would
persist but become arbitrarily heavily damped.
However, when the high energy part has become of order one,
the argument that oscillations should occur with a frequency
$\omega_{osc} \sim \Delta_R$ becomes unreliable and,
in fact, as we have seen above, the conclusion of
the exact solution \cite{TLS})
is that the oscillations 
vanish for $\alpha = 1/2$.

The reason for the success of our perturbation theory,
which is essentially a ``short time expansion'', in
predicting a qualitative change in the tunneling
is that the expansion is valid out to precisely the time when
the spin has order one probability of flipping and is therefore
perfectly adequate to describe the nature of the states reached
by spin flip processes. In particular, it can identify whether
these states
are nearly  degenerate with the initial state
(and each other)
or of widely disparate energies,
which is the essential physical question for
coherence.
Hence, the main conclusions of this section:
the qualitative
behavior of $P(t)$, in the sense of whether or not it exhibits
oscillations,
{\em i.e.\/} quantum coherence, can actually be determined
from lowest order perturbation
theory. The special point $\alpha=1/2$, at which the Golden Rule is
naively applicable, separates a region of completely incoherent
behavior,
$1/2\leq\alpha <1$, from one of damped oscillations, $0<\alpha<1/2$.

\section{The Connection to Fermionic Hopping}
\label{sec:fermions}

The existence of a third regime in the TLS problem
with behavior 
qualitatively different from that occurring for 
irrelevant tunneling or undamped tunneling is suggestive,
however, before we can claim that the lessons of the
TLS have any relevance to the
fermionic hopping 
we must make some firmer connection between the tunneling matrix
element, $\Delta$, and the  single particle hopping,
$t_{\perp}$ between non-Fermi liquids,
in our case one dimensional Luttinger liquids. 

We will study this problem using
bosonization techniques \cite{bosonization,haldjphysc}.
General, gapless, one dimensional interacting electronic systems and 
higher dimensional Fermi liquids can both be studied
via this approach so our results can be made at least
that general.

The Hamiltonians of the isolated systems are of the
general Luttinger liquid form:
\begin{eqnarray}
H & = & 
\frac{1}{4 \pi} \int dx~ \left(
v_{\rho} K_{\rho} (\partial \Theta_{\rho})^2 +
v_{\rho} K_{\rho}^{-1} (\partial \Phi_{\rho})^2 +
v_{\sigma} K_{\sigma}  (\partial \Theta_{\sigma})^2 +
v_{\sigma} K_{\sigma}  (\partial \Phi_{\sigma})^2
\right)
\label{eq:ham_hubb} \\
\nonumber
 & = &  
\frac{1}{4 \pi} \int dx~ \left(
v_{\rho,N}  (\partial \Theta_{\rho})^2 +
v_{\rho,J}  (\partial \Phi_{\rho})^2 +
v_{\sigma,N}   (\partial \Theta_{\sigma})^2 +
v_{\sigma,J}   (\partial \Phi_{\sigma})^2
\right)
\end{eqnarray}
where $K_{\rho}$ is interaction dependent and less than
one for  a repulsive interaction, while $K_{\sigma}$ is
set to one hereafter as a consequence of considering
only interactions which preserve the $SU(2)$ spin invariance.
The bosonized form for the
electron operator is given by
\begin{equation}
\Psi^{\dagger}_{\uparrow}(x) \sim
\sqrt{ \frac{\partial \Phi_{\uparrow}(x)}{ \pi}}
\sum_{m ~odd} A_m ~\exp\left(i [m \Phi_{\uparrow}(x) +
 \Theta_{\uparrow}(x)]\right)
\label{eq:elop}
\end{equation}
so that the  inter-liquid hopping
is given by:
\begin{eqnarray}
\label{eq:hopping}
\Psi^{\dagger,(1)}_{\uparrow}(x) \Psi^{(2)}_{\uparrow}(x)  & \sim &
\sqrt{ \frac{\partial \Phi^{(1)}_{\uparrow}(x)}{ \pi}}
\sum_{m ~odd} A_m ~\exp\left(i [m \Phi^{(1)}_{\uparrow}(x) +
 \Theta^{(1)}_{\uparrow}(x)]\right) \\
\nonumber & \times  & 
\sqrt{ \frac{\partial \Phi^{(2)}_{\uparrow}(x)}{ \pi}}
\sum_{m ~odd} A_m^{\star} ~\exp\left(i [m \Phi^{(2)}_{\uparrow}(x) -
 \Theta^{(2)}_{\uparrow}(x)]\right)
\end{eqnarray} 
where 
\begin{equation}
\Theta_{\uparrow} = 2^{-\frac{1}{2}} (\Theta_{\rho} + \Theta_{\sigma})
\end{equation}
\begin{equation}
\Theta_{\downarrow} = 2^{-\frac{1}{2}} (\Theta_{\rho} - \Theta_{\sigma})
\end{equation}
and
\begin{equation}
\Theta_{\rho}(x) = \Theta_{\rho}^0 + N_{\rho} x/L -i \sum _{q \neq 0} 
|\frac{2 \pi}{q L}|^{\frac{1}{2}} K_{\rho}^{-\frac{1}{2}} ~{\rm sgn}(q)
e^{i q x} \left( b^{\dagger}_{\rho}(q) + b_{\rho}(-q) \right)
\end{equation}
\begin{equation}
\Phi_{\rho}(x) = \Phi_{\rho}^0 + J_{\rho} x/L -i \sum _{q \neq 0} 
|\frac{2 \pi}{q L}|^{\frac{1}{2}} K_{\rho}^{\frac{1}{2}} ~{\rm sgn}(q)
e^{i q x} \left( b^{\dagger}_{\rho}(q) - b_{\rho}(-q) \right)
\end{equation}
where the $b_{\rho}$ operators create and annihilate the bosonic,
charge density eigenexcitations.
Similar expressions obviously apply for $\Theta_{\sigma}$ and $\Phi_{\sigma}$.
The expression for
hopping of down spin electrons is easily obtained by changing
the sign of $\Theta_{\sigma}$ and $\Phi_{\sigma}$
in Eq. \ref{eq:hopping}, while that for
hops in the other
direction can be obtained by interchanging the chain labels in
\ref{eq:hopping}.

In the above expressions the operators $\Theta^0_{\uparrow}$
and $\Phi^0_{\uparrow}$
are canonically conjugate to the the conserved quantum numbers
$J_{\uparrow}$ and $N_{\uparrow}$ and are not expressible in 
terms of the bosons.  The role of these operators was stressed
by Haldane in his solution of the Luttinger model \cite{haldjphysc},
however they are generally ignored since they do not enter into
single particle correlation functions. They will be crucial for
our discussion since it is the quantum numbers 
$N$ and $J$ that are analagous to $\sigma_z$ in the TLS problem.
This is readily apparent when the canonically transformed
form for the tunneling matrix element,
$\frac{\Delta}{2} \sigma^{+} e^{-i \Omega} + ~h.c.$
(see Eq. \ref{eq:TLS_ham_can}), is compared to the bosonized form
for the interchain hopping in Eq. \ref{eq:hopping}.  Both
contain operators which act to raise and lower otherwise 
conserved quantum numbers ($\sigma_z$ for the TLS and 
$N_{\uparrow,1}-N_{\uparrow,2}$, $J_{\uparrow,1}-J_{\uparrow,2}$,
etc. for the fermion hopping).  In addition to the raising
and lowering operators, both contain exponentials in bosonic
creation and annihilation operators which are responsible for
the interesting dynamics and determine the correlation functions
of the operators.  In fact, if the fermionic hopping occurred at only
a single point in space, that problem could be mapped onto
the TLS problem.  We do not believe that such a mapping exists
for the physical model of a uniform interchain hopping.
However, the formal connection
between the models is quite strong and suggests that the interesting
incoherent regime for the TLS might well have a fermionic analogue.

The analogue to $P(t)$ for the fermionic problem
is clear:
instead of taking a system adapted to 
$\Delta =0$ with $\sigma_z = 1$ and then turning on $\Delta$ suddenly,
we take the groundstate of a system with some non-zero values for
$N_{\uparrow,1}-N_{\uparrow,2}$,
$J_{\uparrow,1}-J_{\uparrow,2}$, etc. to $t_{\perp}= 0$ and then turn on
$t_{\perp}$ suddenly.  Instead of studying the resulting oscillations
(or lack thereof)
in $\sigma_z(t)$ we study them in $N_{\uparrow,1}(t) - N_{\uparrow,2}(t)$,
etc.  For simplicity we will hereafter consider only the case
where  the initial condition has
$N_{\uparrow,1} - N_{\uparrow,2}=J_{\uparrow,1} - J_{\uparrow,2}=
N_{\downarrow,1} - N_{\downarrow,2}=J_{\downarrow,1} - J_{\downarrow,2}
=  \delta N(t=0) \neq 0 $, i.e. equal numbers of up and down spin electrons
are added at the right Fermi point of one chain. 
We will follow the dynamics of $\langle \delta N(t \neq 0) \rangle$.

This is a somewhat unfamiliar approach to studying $t_{\perp}$ so it is worth
examining the results for the simple case of free fermions.  In that case,
the problem is exactly soluble. The requirement that the two chains 
be prepared in
states adapted to $t_{\perp}= 0 $ and
in which no Tomonaga bosons are excited but in which
$\delta N(t=0) \neq 0 $ is easily satisfied by simply taking
$n_{1,\sigma}(k) =\Theta ( k_F - k+\frac{2 \pi}{L} \delta N)\Theta (k_F + k)$
while
$n_{2,\sigma}(k) =\Theta ( k_F - k )\Theta (k_F + k)$.
Since the free fermion problem is a single particle one
every $k$ is independent and independent oscillations
occur for the $\delta N$ states for which
$n_{1,\sigma}(k) - n_{2,\sigma}(k) \neq 0$.
The exact result for $\langle \delta N(t) \rangle$ is 
$\delta N(t=0) \cos(2 t_{\perp} t)$.  This is exactly analogous to the 
$\alpha=0$ case of the TLS and clearly corresponds to the interchain hopping
being coherent.  Given this coherence,
it is reasonable to expect that the
description based on degenerate perturbation theory
in $t_{\perp}$ and symmetric and antisymmetric
combinations of the fermion operators for the two
chains will be appropriate.

Notice that in this sense
free electrons (and also Fermi liquids
although we have not shown that here) exhibit
a heretofore unremarked on
{\it macroscopic quantum coherence}: the total number 
difference between
two chains (or planes) of free electrons is a macroscopic
variable which would undergo oscillations, 
rather than incoherent
relaxation, if a finite interchain hopping were suddenly turned
on. Viewed in this light it is not surprising that there might exist
states in which this macroscopic variable loses its coherence.
Rather, it is surprising (though undoubtedly correct for all normal 
metals) that macroscopic quantum behavior should 
occur in generic materials.
It is interesting that this macroscopic quantum coherence 
has not previously occasioned some concern in
the theory of interacting electronic systems. 
It turns out that 
the arguments we have found guaranteeing 
the presence of coherent oscillations  in Fermi liquids
rely
in several places  on 
the quasiparticle structure of the Fermi liquid,
which fails totally for interacting
fermions in one spatial dimension.  We therefore
believe that the postulate
of previous works on arrays of chains of
interacting fermions coupled by a single particle hopping
\cite{Bunch} that
the relevance of $t_{\perp}$ 
signals a crossover to a three dimensionally coherent
Fermi liquid is just that: a postulate.  In fact, we
will see that the
extension
to coupled Luttinger liquids  of the tools
we have used for the TLS problem and
coupled Fermi liquids
does not support the conclusion that a relevant $t_{\perp}$
is always a coherent $t_{\perp}$.

To begin our analysis of coupled Luttinger liquids,
we require the spectral function of the
single particle hopping operator between otherwise
isolated Luttinger liquids.  This is easily obtained 
from the spectral function of the single particle Green's
function.  The universal features of this function
are readily accessible \cite{spectral}.
At the level
of a linearized dispersion relation, the annihilation operator
for momentum $k$ 
has the same spectral function (when the Fermi energy 
contribution to the energies is taken out) as the 
creation operator for momentum $2 k_F -k$.  The hopping operator's
spectral function can be obtained by convolving the
spectral function of the individual creation and annihilation
operators.  

For $\delta N(t=0) = 0$, the spectral function
for $\sum_k c^{\dagger}_1(k) c_2(k)$ is given by
$L ~\alpha \omega^{4 \alpha} \Lambda^{-(1+4 \alpha)}$
where $2 \alpha = \frac{1}{4}(K_{\rho} + K_{\rho}^{-1} - 2)$ 
is the anomalous exponent of the 
single particle Green's function for the case with
spin and $2 \alpha = \frac{1}{2}(K_{\rho} + K_{\rho}^{-1} - 2)$
for the spinless case.
Since the spectral function
vanishes as $\omega \rightarrow 0$ the response to 
$t_{\perp}$ is always incoherent for $\delta N(t=0) = 0$.
This should not be surprising since for $\delta N(t=0) = 0$
there is no possibility of coherent oscillations in 
$\langle \delta N(t) \rangle$.  It is for this reason
that for free particles there
would be no response for $\delta N(t=0) = 0$
since their response is
entirely coherent.  Fermi liquids would have a response
but there would be no long time singular behavior,
the single
quasiparticle hopping having been  completely blocked for
$\delta N(t=0) = 0$.  

Notice that for the Luttinger liquid,
the long time incoherent response {\it is singular}
provided that $4 \alpha < 1$,
despite the fact that coherent hopping is 
totally blocked.  This suggests that 
in some sense
incoherent hopping 
is relevant and that flows away from
the $t_{\perp} = 0$ fixed point may be dominated by this
relevant operator, rather than the relevant operator
corresponding to coherent interliquid hopping.
If this is the case, then the renormalization group flows
should
end elsewhere than Fermi liquid theory.
This is due not only to the anomalous exponent
of the Luttinger liquid, but also the destruction of the
Fermi surface.  No such effect would be present for
a model with a single particle Green's function
of the form $G(k,\omega) \sim (\omega - v k)^{-1+2 \alpha}$.

$\langle \delta N(t) \rangle$
is the natural quantity to study for coherence effects,
however, we will not be able to go beyond lowest
order in perturbation theory 
(which should still be sufficient for
settling questions of coherence as
we argued in the TLS case)
and here
the behavior of $\langle \delta N(t) \rangle$
can be misleadingly coherent because it
involves the subtraction of the
contributions from the 
hops in different directions.
Clearly, if the hopping in both directions
is incoherent, then any coherence in the
difference is an artifact.
It is therefore useful to consider the quantity
$P(t)$, defined as the probability to remain in the
initial state: 
\begin{eqnarray}
\label{eq:samestate}
P(t) & = & |\langle O | \exp(i \int_0^t dt' H'(t') ) | 0 \rangle|  \\
 & \sim & 1 - 4 t_{\perp}^2
\int d\omega \frac{\sin^2(\omega t/2)}{\omega^2}
\left(   \rho_{1\rightarrow2}(\omega) +
  \rho_{2\rightarrow1}(\omega)
\right) + \dots
\nonumber
\end{eqnarray}
Note that oscillations in $\delta N$ are the natural signature
of coherence and no oscillatory behavior in $P(t)$ is expected
in general;
however it is useful for the above
reasons to now restrict ourselves to 
$P(t)$.

Let us now procede with a short
time expansion analogous to that which we used for the TLS.
This should be valid for determining the presence or absence
of coherence for exactly the same reason as it was for that
problem: the presence or absence of coherence is equivalent
to the near degeneracy or non-degeneracy of the states
connected to the initial state by $t_{\perp}$. The short
time expansion is capable of revealing such features
since it is valid out to precisely the timescale 
where  the  initial state has been left behind.

For spinless fermions and
finite $\delta N(t=0) $ ($k_F^1 > k_F^2$)
the initial spectral
function for $\sum_k c^{\dagger}_1(k) c_2(k)$ is
given by:
\begin{eqnarray}
\rho_{2\rightarrow1} (\omega)  & = & 
\Gamma^{-1}(2\alpha) \Gamma^{-1}(2+2\alpha) (2v_S)^{-(1+4\alpha)}
\Theta\left(\omega - (E_F^1 -E_F^2) -v_S(k_F^1-k_F^2) \right) \\
\nonumber &  &
\left(\omega - (E_F^1 -E_F^2) -v_S(k_F^1-k_F^2) \right)^{1+2\alpha} 
\left(\omega - (E_F^1 -E_F^2) +v_S(k_F^1-k_F^2) \right)^{-1+2\alpha} 
\end{eqnarray}

\noindent
likewise

\begin{eqnarray}
\rho_{1\rightarrow2} (\omega)  & = & 
\Gamma^{-1}(2\alpha) \Gamma^{-1}(2+2\alpha) (2v_S)^{-(1+4\alpha)}
\Theta\left(\omega - (E_F^1 +E_F^2)  -v_S(k_F^1-k_F^2) \right) \\
\nonumber &  &
\left(\omega - (E_F^2 -E_F^1) -v_S(k_F^1-k_F^2) \right)^{-1+2\alpha} 
\left(\omega - (E_F^2 -E_F^1) +v_S(k_F^1-k_F^2) \right)^{1+2\alpha} 
\end{eqnarray}

$P(t)$ is radically different, even at
small $\alpha$ from the Fermi liquids case.
In fact for small $\delta N(t=0)$,
the incoherent part of the spectral function
(high energy)
completely dominates
$P(t)$  and 
may  destroy coherence completely,
if the incoherent (high energ) hops effect the coherent
(low energy) ones.
One way to disentangle the coherent
and incoherent competition 
is to consider the spectral functions
broken down into the contributions coming from
individual momenta.
First, examine the spectral function
for $c^{\dagger}_1(k) c_2(k)$, which may be obtained
by convolving the spectral functions for
$c^{\dagger}_1(k)$ and $c_2(k)$.  
The spectral function
for $c^{\dagger}_1(k)$ has support for 
$\omega > E_F^1+ v_S |k - k_F^1|$,
where $v_S$ is the sound velocity of the Luttinger liquid.
The spectral function  behaves at large
$\omega$ like 
\begin{equation}
\rho_{\dagger,1}(\omega ~ large) \sim \omega^{-1+2\alpha}
\end{equation}
and behaves for $\omega \rightarrow E_F^1+ v_S |k - k_F^1|$
like 
\begin{equation}
\rho_{\dagger,1}(\omega ~ small) \sim
\left(\omega - (E_F^1+ v_S |k - k_F^1|) \right)^{\alpha - H(k-k_F^1)}
\end{equation}
where $H(x) = 1$ if $x \ge 0$ and $O$ if $x \le 0$. 
The integrated weight is $1-n_1(k)$.  
The spectral function for
$c_2(k)$ has support for $\omega > -E_F^2+ v_S |k - k_F^2|$,
also behaves at large
$\omega$ like 
\begin{equation}
\rho_2(\omega ~ large) \sim \omega^{-1+2\alpha}
\end{equation}
and behaves for $\omega \rightarrow ~-E_F^2+ v_S |k - k_F^2|$
like 
\begin{equation}
\rho_2(\omega ~ small) \sim
\left(\omega - (v_S |k - k_F^2| -E_F^2)\right)^{\alpha - H(k_F^2-k)}
\end{equation}
The integrated weight is $n_2(k)$.  

The convolution has support for
$\omega > E_F^1 - E_F^2 + v_S |k - k_F^1| + v_S |k - k_F^2|$
which means that except for $k_F^2 \le k \le k_F^1$ the
threshold is $E_F^1 - E_F^2 + 2 v_S |k - k_F^{avg}|$
where $k_F^{avg} = (k_F^1 +  k_F^2)/2$.
The behavior as 
$\omega \rightarrow E_F^1 - E_F^2 + 2 v_S |k - k_F^{avg}|$
is 
\begin{equation}
\rho_{\dagger,1,2}(\omega ~ small) \sim
\left(\omega - E_F^1 - E_F^2 + 2 v_S |k - k_F^{avg}|\right)^{4 \alpha}
\end{equation}

For the case, where $k_F^2 \le k \le k_F^1$, 
the threshold is smallest; there the
$k$ independent threshold is given by
\begin{eqnarray}
\label{eq:wmin}
\omega_{min}(k) & = &
E_F^1 - E_F^2 + v_S (k - k_F^2) - v_S (k - k_F^1) \\
\nonumber
 & = & E_F^1 - E_F^2 +v_S(k_F^1 - k_F^2) \\
\nonumber
 & > & 0 
\end{eqnarray}
The behavior of the spectral function as 
$\omega \rightarrow E_F^1 - E_F^2 + v_S(k_F^1 - k_F^2)$
is 
\begin{equation}
\rho_{\dagger,1,2}(\omega ~ small) \sim
\left(\omega - E_F^1 - E_F^2 + v_S(k_F^1 - k_F^2)\right)^{1 +4 \alpha}
\end{equation}
The positivity of the minimum energy
results  and the large exponent with which
the spectral function vanishes result
from the fact that hops in this direction
are `wrong way' hops, that is they increase
rather than decrease the initial $\delta N$.

The behavior for large $\omega$
is given by
\begin{equation}
\rho_{\dagger,1,2}(\omega ~ large) \sim
\omega^{-1+4\alpha}
\end{equation}
The integrated weight is $(1-n_1(k)) n_2(k)$.  

The spectral function for $c^{\dagger}_2(k) c_1(k)$
is similar.  The spectral function
for $c^{\dagger}_2(k)$ has support for
$\omega > E_F^2+ v_S |k - k_F^2|$ behaves at large
$\omega$ like
\begin{equation}
\rho_{\dagger,2}(\omega ~ large) \sim
\omega^{-1+2\alpha}
\end{equation}
and behaves for $\omega \rightarrow E_F^2+ v_S |k - k_F^2|$
like
\begin{equation}
\rho_{\dagger,2}(\omega ~ small) \sim
\left(\omega - (E_F^2+ v_S |k - k_F^2|) \right)^{\alpha - H(k-k_F^2)}
\end{equation}
The integrated weight is
$1-n_2(k)$. 
The spectral function for
$c_1(k)$ has support for $\omega > -E_F^1+ v_S |k - k_F^1|$,
also behaves at large
$\omega$ like
\begin{equation}
\rho_1(\omega ~ large) \sim
\omega ^{-1+2\alpha}
\end{equation}
and behaves for $\omega \rightarrow ~-E_F^1+ v_S |k - k_F^1|$
like
\begin{equation}
\rho_1(\omega ~ small) \sim
\left(\omega - (v_S |k - k_F^1| -E_F^1)\right)^{\alpha - H(k_F^1-k)}
\end{equation}
The integrated weight is
$n_1(k)$. 

The convolution has support for
$\omega > E_F^2 - E_F^1 + v_S |k - k_F^1| + v_S |k - k_F^2|$
which means that except for $k_F^2 \le k \le k_F^1$ the
threshold is $E_F^2 - E_F^1 + 2 v_S |k - k_F^{avg}|$.
The behavior as
$\omega \rightarrow E_F^2 - E_F^1 + 2 v_S |k - k_F^{avg}|$
is
\begin{equation}
\rho_{\dagger,2,1}(\omega ~ small) \sim
\left(\omega - E_F^2-E_F^1+2 v_S |k - k_F^{avg}|\right)^{4\alpha}
\end{equation}

For the case, $k_F^2 \le k \le k_F^1$
threshold is $E_F^2 - E_F^1 +  v_S (k_F^1 -k_F^2)$,
which vanishes for weak interactions
and is always smaller than 
the threshold for hops in the
other direction.
The behavior as
$\omega \rightarrow E_F^2 - E_F^1 +  v_S (k_F^1 - k_F^2)$
is
\begin{equation}
\label{eq:shiftsing}
\rho(\omega ~ small) \sim
\left(\omega-(E_F^2 - E_F^1 + v_S(k_F^1-k_F^2)) \right)^{-1+2\alpha}
\end{equation}
There is a power law divergence.
The behavior for large $\omega$
is given by
\begin{equation}
\rho_{\dagger,2,1}(\omega ~ large) \sim
\omega^{-1+4\alpha}
\end{equation}
The total weight in the spectral function is given by
$n_1(k)(1-n_2(k))$.

In the region  where
$k_F^2 \le k \le k_F^1$
a Fermi liquid spectral function would
be a delta function at zero frequency, but here
there is a power law singularity at a non-zero,
negative frequency
since $E_F^1 - E_F^2 = (k_F^1 -k_F^2) \frac{v_J + v_N}{2}$
is always larger   than than 
$v_S (k_F^1 - k_F^2) = \sqrt{v_J v_N} (k_F^1 - k_F^2)$ \cite{vJfootnote}.
The essential points are that the singularity is in general
not at zero energy and is a power law rather than a delta
function.

Notice that when $4 \alpha >1$, none of the spectral functions
for the individual momenta 
is decreasing for large $\omega$.  The high energy behavior
is thus incoherent
which implies
that the response in
the $t_{\perp} \rightarrow 0$ limit is
always incoherent at every $k$ at short times.   
The low energy behavior for
$\delta N(t=0) = 0$ is also incoherent. 
This implies that the proposal of
incoherence for $\alpha ^{>}_{-} 1/4$ is self-consistent,
since incoherence leads to a vanishing 
oscillation frequency, $\omega_{\rm osc}$,
and therefore the natural $k_F^1-k_F^2$
to consider is $\omega_{\rm osc}/v_F = 0$, a case
where all the spectral functions are in fact incoherent.
Notice that for
$\alpha < 1/4$ this consistency is not present since
even for $k_F^1-k_F^2 = 0$, some of the spectral
functions associated with momenta near to the Fermi
surface are coherent.
The self-consistency is therefore not trivial
and we believe that there is incoherence for
all $\alpha > 1/4$.
Moreover, we can rule out the possibility that the incoherent phase 
is pushed all
the way to $\alpha = 1/2$, where irrelevance sets in.
This cannot occur because, as $\alpha$ approaches
$1/2$, the 
high energy piece is diverging relative to the low 
energy, just as in the TLS problem.

There  remains the question of the effect of
higher order terms on these arguments. The only
physically significant effect at higher order is
interactions among the hopped fermions which
intuitively should be detrimental to coherence,
pushing it to lower values of $\alpha$ than
$1/4$.  For coherence to be restored at higher
order, the initial spectral functions for the
initial hops would have
to have been gross misrepresentations of 
the spectral functions for later hops.
There is no reason at all for believing that this occurs,
but we have been unable rigorously include 
such high order processes. 

If the fermionic hopping is incoherent and
$\langle \delta N \rangle $ does not oscillate, then the implications
of this for the physics of such a system
should be dramatic. If we follow the usual assumption of
the TLS problem, then the absence of interference
effects in $\langle \delta N \rangle$  implies that
interference effects {\it in general} should not
be observable in such a system for 
histories which involve fermionic hopping between
the Luttinger liquids.  The presence of a higher
dimensional Fermi surface for many chains or
two split Fermi surfaces for two chains are the 
direct result of such interference effects and
one expects that they will not result in the
incoherent regime.  It is for this reason that
the incoherent regime must constitute a totally
different fixed point from states where the
coherence is not confined to the $ab$
plane - the shape of the Fermi surface provides
an order parameter which distinguishes the
two phases in the low energy, long time limit. 
All of these qualitative features can
clearly be carried over to a higher dimensional 
analog where non-Fermi liquid chains are
coupled incoherently.

In this case,
there should be a host of strange physical properties
associated with the incoherence.
The
temperature dependence of the conductivity
is a difficult problem from this
point of view and it is not clear that the
activated behavior seen in underdoped
YBa$_2$Cu$_3$O$_{6+x}$ and Bi$_2$Sr$_2$CaCu$_2$O$_8$
can be explained by incoherence alone.  It more
likely results from the
presence of some other relevant operator 
such as an interplane $J$.
However, the incoherence proposal seems
to work remarkably well for the
other experimental results.
For instance,
we expect an optical
conductivity in the $c$ direction which exhibits no
Drude peak at low frequencies and rises weakly with
frequency at high frequencies \cite{prl2}.
The idea of confined coherence is at least consistent with all
of the know behaviors of $c$ axis transport in 
the cuprates, and, in particular, it is consistent
with the peculiarly contradictory features
discussed in Section \ref{sec:motiv}.  
However, the greatest experimental support
for the proposal comes from experiments on
an anisotropic organic conductor,
(TMTSF)$_2$PF$_6$.  As we will now discuss,
this material directly exhibits a low
temperature phase in which interference
effects between histories involving
interplane hops are totally absent
\cite{magic}!

\section{Incoherence in (TMTSF)$_2$PF$_6$}

(TMTSF)$_2$PF$_6$  is a highly anisotropic
organic conductor.  
The material is triclinic and the
three real space lattice directions
are denoted by $a$, $b$ and $c$, respectively,
with 
the hopping integrals
in the three different directions 
estimated to be approximately $250 ~meV$,$25 ~meV$ and $1 ~meV$.
The Coulomb energy to put two electrons
in the same unit cell is estimated at about
$1eV$, so that the material should be strongly
correlated as well as highly anisotropic.
This is especially true in light of
the fact that the conduction band of the 
material is half-filled at low temperatures
due to a dimerization and the fact that
the anisotropy is so large that the corresponding
Fermi surface should be topologically one
dimensional, consisting of a pair of warped 
Fermi sheets.  

\begin{figure}
\centerline{ \epsfxsize = 5.0in
\epsffile{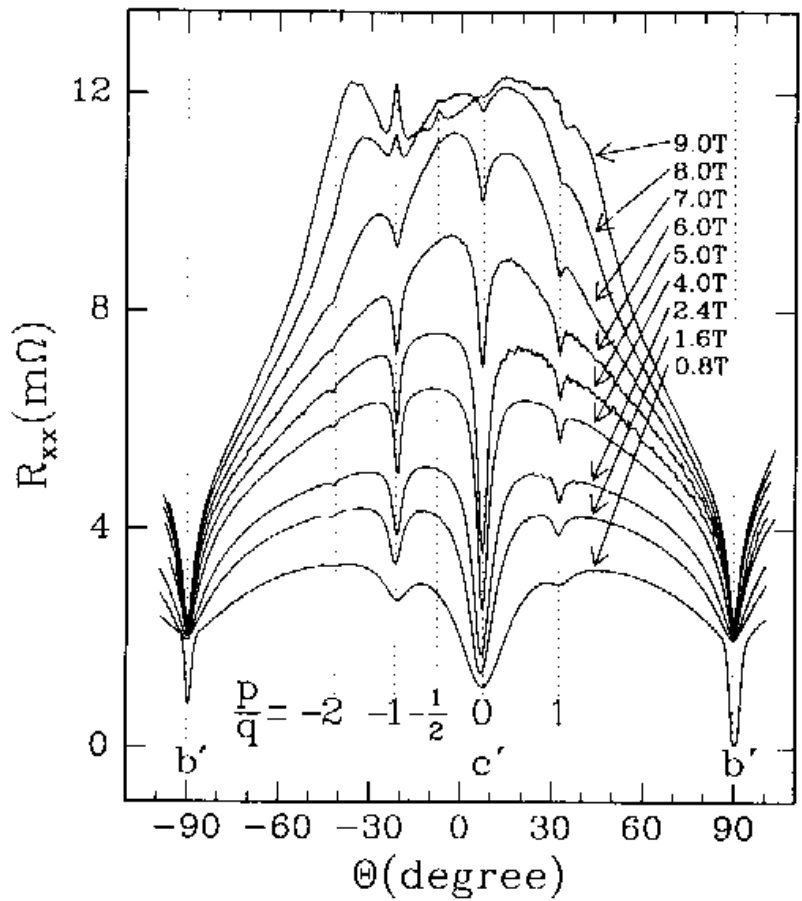}}
\caption{Resistance along the most conducting direction
(in milliohms) as a function of magnetic field strength
and orientation.  Field was rotated in the $bc$ plane
and angle $\Theta$ is defined so that
$\pm$90 degrees  coincide with the $b$ direction.
Data are
taken from Ref. 21}
\label{fig:woowon}
\end{figure}

Reflecting the strong correlation effects,
the low temperature phase diagram of
the material is quite rich: the material 
is in a spin density wave state at
ambient pressure and temperatures
below about 12 Kelvin, but becomes either
superconducting (below about 1K)  or metallic
(at higher temperatures) on the application of pressure.  
In addition, in high magnetic fields and pressures
there is a cascade of field induced spin density
wave transitions \cite{fisdwmagiclebed}.  For our purposes,
we are interested in the metallic phase that occurs above 
about 6 kilobars of pressure and above 1 Kelvin
(in zero field)
or at lower temperatures in applied magnetic fields
of a few Tesla.  In this region the material
exhibits highly anomalous magnetoresistance
behavior \cite{woowon}
(see Figure \ref{fig:woowon})
Specifically, for the resistivity
measured either in the most conducting direction,
parallel to $a$, or in the least conducting
direction, perpendicular to the $ab$ planes,
resistivity is a strong function of magnetic field
strength and orientation.  For a material with
isotropic scattering and a topologically one
dimensional Fermi surface, the magnetoresistance
in the most conducting direction
vanishes identically, so the greater than fivefold increases
possible in the resistivity are highly
surprising.  It is not believed that the effect can be 
accounted for by an anisotropic scattering
rate with any reasonable assumptions about the
electron-electron interaction \cite{yakovenko}.

In addition to the anomalously
large magnitude of the magnetoresistance,
there is also a striking angular dependence
with sharp minima occurring whenever the magnetic
field parallels a real space lattice vector. These
features are referred to as ``Lebed magic angles'',
following a suggestion of Lebed's that
there should be features in the fields induced spin
density wave transition for fields with these 
orientations \cite{fisdwmagiclebed}.  There are a number of 
theoretical proposals to account for the presence of
these commensurability features \cite{magicothers},
however, we will see that a magnetic field
induced confinement of coherence to the $ab$ 
planes will provide a very natural one,
and one which leads to a number of strong,
and strikingly confirmed,
predictions.

The scenario we envision is the following: due to
the strong correlation and large anisotropy, conduction
in the $c$ direction in (TMTSF)$_2$PF$_6$ is
nearly incoherent and, in this sense, the material
is nearly two dimensional.  In this case it is
important that an applied magnetic
field with a component  perpendicular to the
$ac$ plane will 
render hopping in the $c$ direction somewhat
inelastic.  To see this inelasticity,
one encorporates the 
field via a Peierls substitution,
in which case,
one finds that the states connected by
the hopping no long have the same momentum 
in the $a$ directions but rather differ in
momentum by $\frac{e l_c H_b}{c}$, and since
there is a well defined $v_F^a$ due to the
topologically one dimensional Fermi surface,
the states connected by hopping in the $c$ direction
will have their degeneracy lifted by
$\frac{e l_c H_b v_F^a}{c} \sim H_b \cdot 0.2 meV/Tesla$.
This added inelasticity will enhance the
non-degeneracy of the perturbation theory in
$t_c$, the hopping in the $c$ direction, and
if conduction were nearly incoherent before,
the field may drive the material into the purely incoherent
regime.  This effect clearly won't occur if
the magnetic field is purely in the $c$
direction, and one expects the 
resistivity to dip sharply there.
The out of $ab$ plane  resistivity
should have the strongest dip near
$c$ since it is precisely the conduction
in that direction that is changing character
at the transition;
however,
the transition should be also be associated with a large
change in the $a$ resistivity, since the material
is really changing between two different states,
and the scattering effects in the incoherent
state ought to be more pronounced due to an effectively
reduced dimensionality.  This can clearly explain the
angle dependence of the magnetoresistance around 
$c$ and, for any substantial hopping integrals
in the $\hat{c} \pm \hat{b}$ direction, the 
dips associated with those directions as well,
however, the real strength of the theory lies
in the other predictions that it leads to.

One immediate prediction 
is that the minimum in the magnetoresistance
associated with fields parallel to $b$
has a different origin than the other minima:
it is not associated with the restoration
of coherence and is therefore not
a magic angle effect.  This 
difference is readily apparent in the data
for resistivity in the $a$ direction
where the magnetoresistance has a cusp
like behavior at $b$. Moreover,
the value of the magnetoresistance for
fields in the $b$ direction is field
independent above 1 Tesla, while this is
not true of the other minima. 
It is important to note that the change in
resistivity out of the $ab$ planes
is only order one for large fields directed along
$b$.  If the magnetic fields were somehow
rendering the interplane hopping irrelevant
then we would not expect  the
a much larger change in the
resistivity
than this at temperatures of
0.5 Kelvin and we would not 
expect any saturation of the
magnetoresistance.  
Likewise a Fermi liquid explanation of
the resistivity  out of the $ab$ planes
predicts no saturation of the resistivity for
this field orientation.
The observed behavior
really only makes sense if there 
is a large incoherent hopping that is
relatively unaffected by the magnetic
field and a small coherent hopping that is
wiped out by the field.

\begin{figure}
\centerline{ \epsfxsize = 5.0in
\epsffile{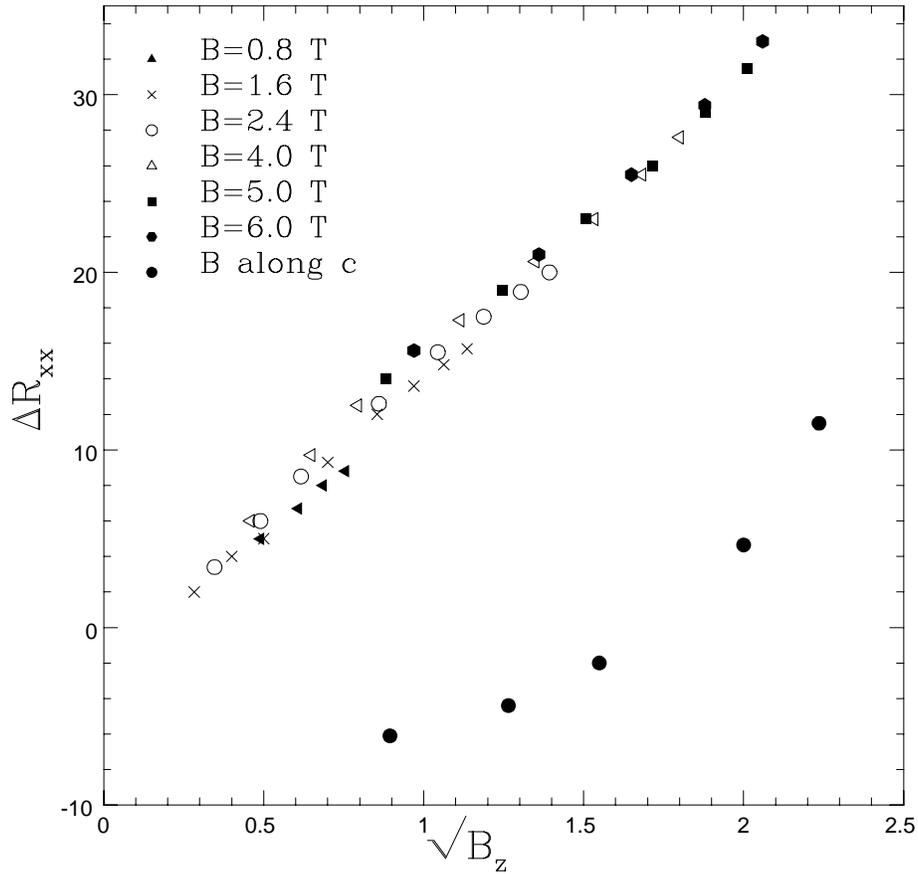}}
\caption{Magnetoresistance along the most conducting direction
(in milliohms) as a function of
the square root of the component of the magnetic field
perpendicular to the $ab$ plane.
As explained in the text, the data are expected to
scale away from the magic angle dips.
Data taken from Ref. 21}
\label{fig:scaling}
\end{figure}

Let us now discuss the behavior away from $b$ and
from the magic angles.
The
field strength independence for fields oriented along $b$
is a direct consequence of
the most powerful prediction of the
incoherence explanation of the magic angles
behavior:
the scaling of the magnetoresistance.  
Since the incoherent
regime  is categorized by the impossibility of
observing interference effects between histories
in which particles leave the $ab$ planes, it
is clear that, neglecting Zeeman
effects and treating the magnetic field
again at the level of a Peierls substitution,
all physical properties should be independent of
the magnetic field components that lie in the $ab$
plane.  This is because in a path integral
calculation of any physical quantity,
these components of the fields enter 
through a phase factor proportional
to the flux enclosed by these paths
that leave the $ab$ plane.  This phase,
however, must be unimportant if 
interference effects between these paths
are blocked by the incoherence.
The magnetoresistance in the $a$ direction
should therefore demonstrate a kind of scaling
behavior away from the magic angle dips, so
that if the resistance is replotted versus only
the component of the field out of the $ab$
plane, then the curves obtained by rotation of
fields of different  strengths should all lie
on top of each other.
Such a plot is shown in Figure
\ref{fig:scaling} 
and the extent to which
the results satisfy the prediction is striking.

\begin{figure}
\centerline{ \epsfxsize = 5.0in
\epsffile{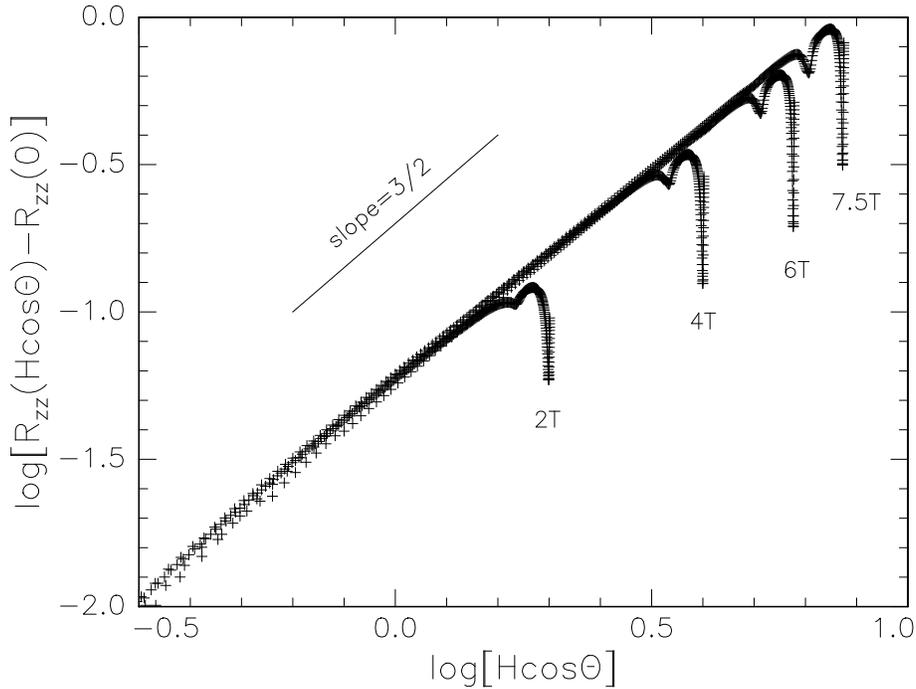}}
\caption{Log of the resistance along the least conducting direction
(in ohms) as a function of
log (base 10) of the component of the magnetic field (in Tesla)
perpendicular to the $ab$ plane.
As explained in the text, the data are expected to
scale away from the magic angle dips.
Raw data were provided by Danner, {\it et al.}.
}
\label{fig:rzzscaling}
\end{figure}

Even more striking is the scaling of the 
resistance out of the $ab$ plane
shown in Figure \ref{fig:rzzscaling},
since in this case the resistance depends only
on the component of the magnetic field {\it parallel}
to the current!   
In both cases, the magnetoresistance
depends on the field through some anomalous
power law: $\Delta R \propto (\vec{H}\cdot\hat{n}_{ab})^x$
with $x \sim 1/2$ for resistivity in the $a$ direction
and $x \sim 3/2$ for resistivity in the $c$ direction.
We are not able to calculate the form
of the scaling functions that our theory predicts
at the present time, however,
the presence of such anomalous power laws is 
broadly compatible
with the picture of incoherently coupled non-Fermi
liquid plans, while
these anomalous powers are not
compatible with any of the previous
proposals to account for the magic angles
behavior \cite{magicothers}.

This scaling should only occur away from the magic angles
and for sufficiently large fields, since otherwise
the material is three dimensionally coherent.  The
points taken from the central magic angle dip in Figure
\ref{fig:scaling} clearly violate the scaling exactly
as we expect, so the prediction is satisfied at that
level. The low field anisotropy in the magnetoresistance
is currently being investigated by Chaikin {\it et al.}
\cite{chaikinunpub}
and the prediction that it should not satisfy
the scaling that occurs in the incoherent regime
will thus be tested in the near future; the present
evidence is that the scaling is violated in exactly the
manner expected at low fields.

Danner, {\it et al.} \cite{guynfl} have also investigated
the coherence question in (TMTSF)$_2$PF$_6$
directly using a novel resonance in the $c$ axis
conductivity discovered by them \cite{batman}.
The resonance is a Fermi liquid effect
associated with the averaging
of the $c$ axis velocity to zero in
a magnetic field due to the quasiclassical
trajectories over the Fermi surface.  It 
should therefore be present in (TMTSF)$_2$PF$_6$
only in the coherent regions.
This is exactly what is  found
by Danner, {\it et al.} experimentally \cite{guynfl}. 

There is thus ample experimental evidence that
(TMTSF)$_2$PF$_6$ undergoes a transition in
magnetic field at low temperatures to a state
in which coherent transport is confined to the
$ab$ planes of the material.  The effect is more
pronounced in cleaner samples and at lower temperatures
(thus ruling out impurity- or thermally-induced incoherence)
and, we believe, provides direct experimental evidence
for the existence of a new fixed point with
interaction induced confinement of the coherence.



\begin{references}
\bibitem{general} The
notion of intrinsic incoherent transport resulting
from interactions is clearly more general than the
model we have considered, and our calculation is really meant
to establish that such incoherence is possible.
More generally, we can expect it to occur
for sufficiently strong
interactions if the system, for $t_{\perp} = 0$,
(1) is gapless,
(2) lacks  the quasiparticle structure  of a Fermi liquid
and (3) lacks a sharp Fermi surface (not necessarily
implied by 2).
\bibitem{vJfootnote} Here we considered a
$\delta N$ at only the right Fermi point,
if we add equal numbers at both Fermi points then
$E_F^1 - E_F^2 = (k_F^1 -k_F^2) v_N$, which is again in
general different from
$v_S (k_F^1 - k_F^2) = \sqrt{v_J v_N} (k_F^1 - k_F^2)$
but may be smaller if the interaction is attractive.
\end{references}
\end{document}